\documentclass[%
 reprint,
 amsmath,amssymb,
 aps,
]{revtex4-2}

\usepackage{graphicx}% Include figure files
\usepackage{dcolumn}% Align table columns on decimal point
\usepackage{bm}% bold math
\begin{document}

\preprint{APS/123-QED}

\title{Energetics of Feedback: Application to Memory Erasure}% Force line breaks with \\
%\thanks{A footnote to the article title}%

\author{Doddi Harish}
\email{doddi003@umn.edu}
\affiliation{University of Minnesota}%
\author{Saurav Talukdar}
\email{taluk005@umn.edu}
\altaffiliation[Also at ]{University of Minnesota}%Lines break
\author{Murti Salapaka}%
\email{murtis@umn.edu}
\altaffiliation[Also at ]{University of Minnesota}%Lines break automatically or can be forced with \\

\date{\today}% It is always \today, today,
             %  but any date may be explicitly specified

\begin{abstract}
% Feedback in a closed-loop system is an action performed by the controller based on a measurement. The connection between information theory and thermodynamics has received significant attention over the past decade. The thermodynamic limits of feedback is harnessed to erase a memory bit in finite time using time multiplexing of Optical Tweezers; with considerable attention given to experimental artifacts, such as imperfections in memory model, irreversibility and imprecise measurements. We propose a closed-loop erasure protocol involving measurement and feedback. Our experimental realization of the average feedback work necessary for successful erasure, under closed-loop protocol, is smaller than the Landauer limit, $k_BT\ln2$. Such deficit in the energy expenditure is accounted by the mutual information between the measurement and the state of the system, in a quasistatic limit. Further, we elucidate the role of measurement noise on the probability of erasure.
Landauer’s erasure principle states that any irreversible erasure protocol of a single bit memory needs work of at least $k_BT \ln 2.$ Recent proof of concept experiments has demonstrated that the erasure protocols with work close to the Landauer limit can be devised. Under feedback, where the state of the bit can be measured, the work needed for a bit erasure can be lower than $k_BT\ln 2$. In this article, we analyze the energetics of feedback enabled erasure, while incorporating the imperfections of experimentally realized memory and bit erasure protocols that admit failure probabilities. We delineate the role of uncertainty in measurements and its effects on the work and entropy changes for a feedback-based erasure. We quantitatively demonstrate that the deficit between the Landauer limit and the minimum average work needed in a feedback-based erasure is accounted for by the mutual information between the measurement and the state of a memory, while incorporating the imperfections inherent in any realization. We experimentally demonstrate analysis results on a memory and erasure protocol realized using optical fields.
\begin{description}
\item[Keywords]
Feedback Erasure, Mutual Information, Measurement Noise
\end{description}
\end{abstract}

%\keywords{Suggested keywords}%Use showkeys class option if keyword
                              %display desired
\maketitle

%\tableofcontents

\vspace{-8pt}
\section{\label{sec:Intr}Introduction}
\vspace{-8pt}
The connections between feedback, measurement and entropy are studied by numerous works; originally, the connection was expounded by Maxwell with a thought experiment of a demon separating fast and slow moving molecules in a fixed container, based on their speed, creating a paradox where the demon was able to create a temperature difference without an external work \cite{leff2014maxwell, rupprecht2019maxwell}. Works by Bennett \cite{bennett1985fundamental} indicated that the demon is playing a role of a controller, which uses a measurement of the molecule's velocity in a feedback action of opening or closing a partition separating the two chambers. An insight reached is that the energetics of a measurement needed for feedback cannot be ignored. 

Recent analytical works addressing thermodynamics and feedback include \cite{sagawa2012nonequilibrium, cao2009thermodynamics,sordal2017influence}, with a focus on non-equilibrium processes with extensive simulation studies \cite{sordal2017influence,dinis2021extracting,abreu2011extracting}. With new capabilities of exploring and controlling systems with energetics at the scale of $k_B T,$ where $k_B$ is the Boltzmann constant and $T$ is the temperature, the interdependencies of feedback and thermodynamic quantities can be explored experimentally with the associated limitations imposed by non-ideal conditions not considered in the analysis. Recent advances in probing and controlling of microscopic systems have enabled researchers to verify the fundamental limits, such as the Landauer limit, experimentally \cite{berut2012experimental, talukdar2017memory}. An experimental demonstration of information to energy conversion assuming feedback with perfect measurements was reported in \cite{toyabe2010experimental}.

In this article, the thermodynamic limits associated with measurement and feedback are analyzed while addressing non-ideal conditions, instantiated for the erasure process. We address the limits on the energy savings possible with feedback while employing imperfect measurements. We illustrate our results experimentally with precise accounting of the link between information and thermodynamics of the erasure process. 

Erasure is a process of resetting the state of a memory bit, with its states $0$ or $1,$ to $0$ (reset state) irrespective of its initial state. Erasure is a fundamental operation in information processing and associated with heat release. Erasure efficiency is at the heart of our ubiquitous dependence on cloud computing, where heat released from the erasure is significant. There is an increased effort to design energy-efficient erasure protocols and operate a datacenter with renewable energy; for example, Microsoft {\it Project Natick} \cite{cutler2017dunking}. Landauer ($1961$) claimed that the average work needed for a successful erasure of a binary memory bit is at least $k_BT\ln2,$ termed as Landauer limit. Landauer assumed a particle in a bistable potential, separated by an infinite barrier height, to be an ideal memory model. We deal with a practical memory model with finite barrier height (see FIG. \ref{fig:plot1}(a)), where, given sufficient time a particle can jump from one well to another with a finite non-zero probability. 

A particle assumes state $0$ or $1$ depending on whether it is located in a left or right well. An erasure can be achieved by modifying the potential to affect the transfer of the state to a desired reset state, before returning to the nominal potential. An open-loop erasure protocol is agnostic to the current state of a particle, whereas, in a closed-loop/feedback erasure protocol, the information on the current state of the memory is utilized to affect the transfer. Performance of an erasure protocol is assessed based on the probability of erasure; we term the erasure protocol to be admissible only if the probability of erasure is greater than 95$\%$. 

Recent experimental verification of the Landauer limit include \cite{berut2012experimental,talukdar2017memory}, that employ open-loop erasure protocols. \cite{talukdar2017memory} demonstrates a proof of concept experiment to attain the Landauer limit. In \cite{talukdar2017memory}, the erasure protocol entails lifting the right well and lowering the left well simultaneously, so that the particle ends up in the left well (reset state) with high probability. In \cite{talukdar2017memory}, no measurement and feedback are employed. As open-loop erasure protocols disregard the state of a memory, the protocol does work even if the state is currently in the reset state. With feedback, there is a possibility of spending less work needed for erasure; here, if the state is measured to be in the reset state, then there is no need to execute the protocol for erasure. However, a faulty measurement with a decision that the state is in the reset state may leave the state unchanged, leading to no erasure and decreasing the probability of erasure; thus, the energetics of the sensor state (used for measurement) and the feedback protocol need a careful analysis.

A feedback erasure protocol involves measurement of the particle position followed by a feedback action. We denote by, $W_{fb}$ and $W_{open}$ to be the work done in a feedback action and open-loop erasure protocol, respectively. We quantify the lowest possible expected value $\langle W_{fb} \rangle$ of work using feedback protocol, and address the limits on the difference between $\langle W_{fb} \rangle$ and $\langle W_{open} \rangle$. Regardless of the physical design of a memory bit, such as in optical or colloidal or biological systems, the following analysis and the conclusions hold true because they are based on the fundamental laws of thermodynamics. We introduce the thermodynamic description of the measurement and feedback system in the following section. 

\vspace{-12pt}
\section{Thermodynamic description of a Feedback system}
\begin{figure}
		\includegraphics[width=1.01\columnwidth]{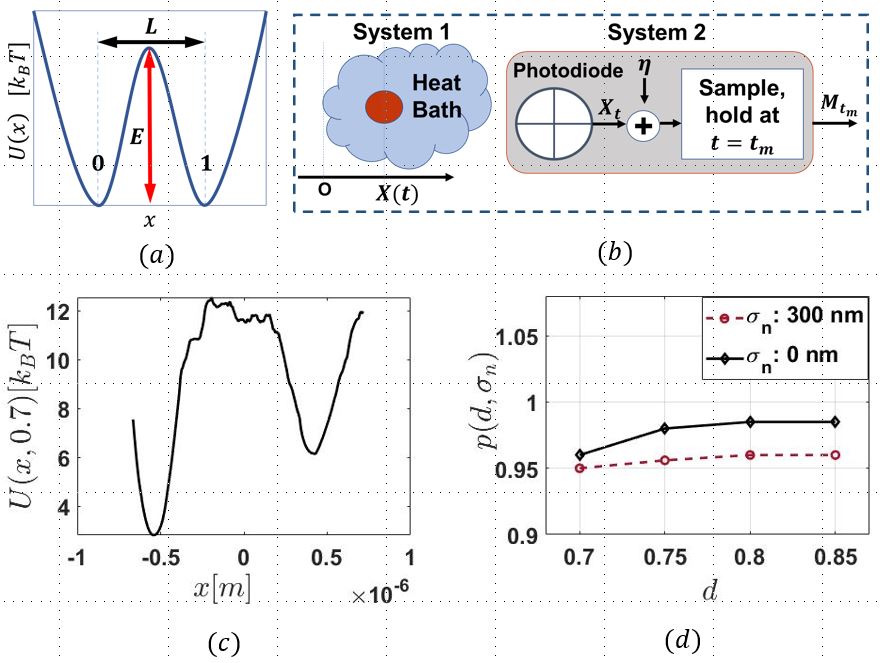}
	\caption{(a) Memory model (b) Thermodynamic system of interest (red  colored circle represents a brownian particle) (c) Asymmetric potential for $d=0.7$ (experiments) (d) Experimental values of $p(d,\sigma_n).$ 	\label{fig:plot1}}
\end{figure}

\vspace{-8pt}
We utilize Optical Tweezers to create $k_BT$ level energy landscape needed for creating a bit memory. The potential wells are manipulated using time multiplexing of lasers \cite{bhaban2016noise}. System $1$ consists of a brownian particle in a thermal environment of temperature $T$. System $2$ consists of a photodiode and a digital sampler, termed as measurement device. Systems $1$ and $2$ together constitute the thermodynamic system of interest and the rest is considered as surrounding. FIG. \ref{fig:plot1}(b) shows the thermodynamic system. Let $X_t$ and $M_t$ denote the true and measured position of a particle at time $t$, respectively. 

% Hamiltonian of the system is described by $\mathcal{H}(X_t,M_t)= \mathcal{H}(X_t)+\mathcal{H}(M_t)+\phi(X_t,M_t),$ where, $\mathcal{H}(X_t)$ and $\mathcal{H}(M_t)$ represents the Hamiltonian's of system $1$ and $2$ respectively and $\phi(X_t,M_t)$ represents the interaction potential between the system $1$ and $2.$ The photodiode in the system $2$ obtains the particle position without affecting the distribution of $X_t$. Hence, $\phi(X_t,M_t)$ is considered negligible. Overall, the Hamiltonian of the system is $\mathcal{H}(X_t)+\mathcal{H}(M_t).$

We model the probability density of $X_t$ via $f_{X_t}(x) = p\mathcal{N}(-L,{\sigma_T}^2)+(1-p)\mathcal{N}(L,{\sigma_T}^2)$, where $\sigma_T$ represents strength of thermal noise, $p$ is the probability of a particle being in the left well, and $\mathcal{N}(L,{\sigma_T}^2)$ is a normal distribution. The joint probability distribution of the system is given by 
$f_{X_t,M_t}(x,m).$ The entropy of the system is determined by evaluating Shanon/information entropy of  $f_{X_t,M_t}(x,m)$. The mutual information between $X_t$ and $M_t$ is $I(X_t,M_t)= $ $\int \int f_{X_t,M_t}(x,m) \ln(\frac{f_{X_t,M_t}(x,m)}{f_{X_t}(x)f_{M_t}(m)})\ dxdm.$ Let $\mathcal{F}(X_t,M_t)$ denote the non-equilibrium/information free energy of the system. The difference between the information free energy and the Helmholtz free energy, $F(X_t,M_t),$ is directly proportional to the Kullback–Leibler divergence between $f_{X_t,M_t}(x,m)$ and the canonical distribution of the system, $f_{eq}(x,m)$ ; that is, $\mathcal{F}(X_t,M_t) = F(X_t,M_t) + k_BT \int \int f_{X_t,M_t}(x,m) \ln(\frac{f_{X_t,M_t}(x,m)}{f_{eq}(x,m)})\ dx dm$. See \cite{deffner2012information} for more details. The non-equilibrium free energy of the system is given by $\mathcal{F}(X_t,M_t):= \mathcal{F}(X_t)+\mathcal{F}(M_t)+k_B T I(X_t;M_t)$. We now proceed with the feedback erasure and its thermodynamics.
\vspace{-11pt}
\section{Feedback Erasure}
\vspace{-8pt}
We propose a feedback erasure protocol for erasing a memory bit in time $t_e$. The protocol involves three steps:

(a) Measurement: At $t=0,$ the measurement device is assumed to be at nominal state $0.$ At $t=t_m$, a measurement of the particle position is obtained, with the state of the measurement device changing to $M_{t_m}$. The state of the system of interest before and after the measurement are $\{ X_{t_m}^{-},M_{t_m}^{-}=0 \}$ and $\{ X_{t_m},M_{t_m}\},$ respectively. The random variables $M_{t_m}$ and $X_{t_m}$ are dependent, whereas $M_{t_m}^{-}$ and $X_{t_m}^{-}$ are independent.

(b) Feedback: If $M_{t_m}$ is strictly positive, then the right well is raised and the left well is lowered simultaneously, facilitating the particle transport to the left well (reset state) with high probability. The system is left with the asymmetric potential until time $t_f:=(t_m+\tau)$ sec, after which the potential wells are restored to initial symmetric configuration. On the other hand, if $M_{t_m}$ is negative, then there is no action taken because the memory is already in the reset state; then, the external work done is zero as the potential wells are unaltered. The system is further relaxed until $t_e.$ The value of $\tau$ needed for successful erasure depends on the extent of the asymmetry of the potential wells, if manipulated \cite{talukdar2017memory}. The feedback action is assumed to not affect the state of the measurement device and hence $M_t$ is same as $M_{t_m}$ for $t_m \leq t < t_e.$ 

(c) Reset: At time $t_e$, the value stored in the measurement device is now reset to zero without affecting the distribution of $X_t.$

We now describe an experimental setup to create a memory model and perform the feedback erasure protocol.
\vspace{-8pt}
\subsection{Feedback Erasure Experimental Setup}
\vspace{-5pt}

A sample of polystyrene particles suspended in an aqueous solution (heat bath) is placed in an Optical Tweezer. A particle is trapped by a laser (see \cite{bhaban2016noise}) near the focus of the objective lens with high numerical aperture; thus creating a single harmonic potential $U(x)$ around the trap center, where $x$ is the particle position. The model for the single well potential is,
\small
\begin{align} \label{eqn:single_well}
  U(x) = \begin{cases} \frac{1}{2}kx^2+U_r, &\mbox{if } |x| \leq w,\\
   \frac{1}{2}kw^2+U_r, & \mbox{otherwise, } \end{cases}
\end{align}
\normalsize where $U_r$ is a constant reference potential energy, $k$ is the stiffness and $w$ is the width. The particle position is measured using a photodiode of $1nm$ accuracy at a bandwidth of $20 kHz.$ The position data $x$ is collected for sufficient time for accurate estimation of probability distribution $P(x)$. With a particle in thermal equilibrium with the heat bath, the particle position is given by Boltzmann distribution; solving for $U(x)$ gives $-k_BT\ln (\frac{P(x)}{C})$. Experimentally computed values of $w$ and $k$ for a single potential well are $175 nm$ and $0.0045\ pN/nm,$ respectively. Refer to \cite{bhaban2016noise, talukdar2017memory} for more details on experimental setup.  

For our memory model, we require a double well potential with two stable equilibrium locations at $L$ and $-L$. This is achieved by alternately focusing a trapping laser at two trap locations using time multiplexing enabled by an Acousto Optical Deflector. The time spent by the laser at a trap location ($\sim 10 \mu s$) is much smaller than the time constant of the particle ($1\ \sim ms$), which helps in creating a bistable potential well. Duty ratio ($d$) is defined as the fraction of the total time spent by the laser at left location. Duty ratio provides a control on the asymmetry of the bistable potential. For $d=0.5$, the two wells are symmetric; for a duty ratio of $0.7,$ the wells are asymmetric, as shown in FIG. \ref{fig:plot1}(c). 

For creating a stable memory bit (where the time of exit from one state to another is considerably longer than the experiment time), $L$ is chosen to be $550 nm$. A particle is initialized either in the left or right well with equal probability and allowed to thermalize with the surrounding heat bath. The mathematical model of the bistable potential $U(x,d)$ is characterized by the model of the single well potential in Eq. (\ref{eqn:single_well}) as follows:
\small
\begin{align}\label{eqn:twoPotential}
  U(x,d) = \begin{cases} \frac{1}{2}k(x-L)^2+U_r, &\mbox{if } |x-L| \leq w,\ r(t)=1,\\
  \frac{1}{2}k(x+L)^2+U_r, &\mbox{if } |x+L| \leq w,\ r(t)=0,\\
\frac{1}{2}kw^2+U_r, & \mbox{otherwise, }
\end{cases}
\end{align}
\normalsize
where, $r(t)=1$ when the laser is present at right trap, otherwise $r(t)=0$; $w,\ k$ are the parameters that we computed for the single potential well earlier. The particle dynamics under the influence of time multiplexed potential is modeled by overdamped Langevin equation,
\small
\begin{align}\label{eqn:langevin_multiplexed}
  -\gamma \frac{dx}{dt}+\xi(t)-\frac{\partial U(x,d)}{\partial x}=0.
\end{align}
\normalsize
Here, $\xi$ is a zero-mean uncorrelated Gaussian noise with $\langle \xi(t),\xi(t') \rangle = 2D \delta(t-t'),$ where $D= \gamma k_BT$ is the diffusion constant and $\gamma$ is the coefficient of viscosity. See \cite{bhaban2016noise} for more details. Monte Carlo simulations are performed using Eq. (\ref{eqn:langevin_multiplexed}) in conjunction with Eq. (\ref{eqn:twoPotential}) for a duty ratio of $0.5$. From the ensembles of $\{x(t) \}$ simulated, the potential wells are reconstructed using the canonical distribution. The bistable potential reconstructed using simulated data matched closely with the potential wells constructed from the position data collected from photodiode. The measurement device consists of sensor and a time sampler as shown in FIG. \ref{fig:plot1}(b). The sensor is composed of a photodiode signal corrupted with a measurement noise $\eta .$ The measurement signal $M_{t_m}$ is modeled as $X_{t_m}+\eta$, with $\eta \sim \mathcal{N}(0,\sigma_n)$ is the measurement noise. 

Using the memory model and the measurement device, the feedback erasure protocol is implemented. Experimentally it was ascertained that the protocol achieves erasure with a probability $> 95 \%$. As alluded earlier in the Feedback step of the protocol, the potential wells are made asymmetric for successful transfer of particle to the left well. Duty ratio $d$ controls the asymmetric nature of the wells during the feedback step. For each $d$ in $\{0.65,0.7,0.75,0.8,0.85\},$ we instantiated $300$ independent runs of erasure based on feedback. For $d=0.65,$ the probability of the successful erasure under open-loop protocol is less than $95 \%$; the results for $d= \{0.7,0.75,0.8,0.85 \},$ where the open-loop protocol results in erasure with a probability higher than $0.95$ are reported.

\vspace{-12pt}
 \subsection{Probability of Successful Erasure}
 \vspace{-5pt}
The photodiode measurement $M_{t_m}$ is a continuous random variable, modeled as $X_{t_m}+\eta,$ where, $\eta \sim \mathcal{N}(0,\sigma_n)$ is measurement noise. The conditional density $f_{M_t/X_t}(m,x) = f_{\eta}(m-x),$ where $f_{\eta}$ is the density function of a zero-mean Gaussian with variance $\sigma_n^2$. In an open-loop protocol \cite{talukdar2017memory}, the probability of successful erasure depends on the duty ratio. A duty ratio greater than $0.7$ results in $p>0.95.$ However, in a feedback erasure protocol the value of $p$ depends on both $d$ and $\sigma_n.$ Because, the measurement may wrongly indicate that the particle is in the well corresponding to the reset state. Such a measurement will lead to no action or reset which will lead to an unsuccessful erasure. Moreover, the success of erasure has considerable impact on the minimal amount to work needed for an erasure. 
\begin{figure}
	\includegraphics[width=0.85\columnwidth]{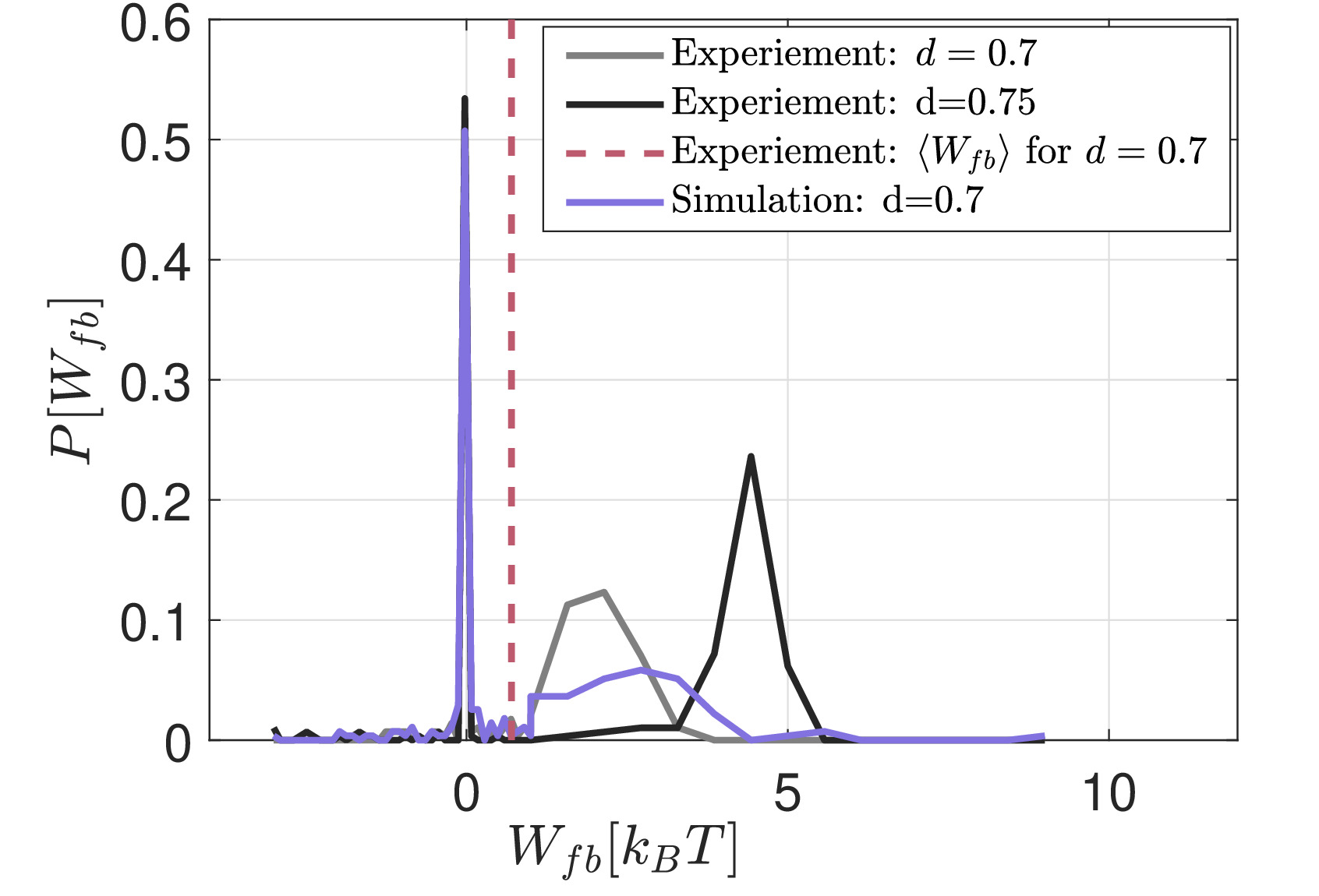}\\
\caption{Distribution of $ W_{fb}^d $: spike at $0$ is for $0-0$ erasures.}\label{fig:pdf}
\end{figure}
An estimate on the work needed to erase a single bit memory with a success probability $p$ is provided by the Generalized Landauer bound (GLB), $k_BT[\ln 2+p\ln p + (1-p)\ln (1-p)]$ \cite{talukdar2017beating}. Here, the erasure probability $p>0.95$ ensures that GLB is within $29 \%$ of the Landauer limit. As discussed earlier in the feedback erasure protocol, the erasure probability $p$ depends on the specifics of how the wells are altered to execute transfer of the state as well as the quality of the sensor that impacts whether the state transfer protocol is executed or not. We now relate the open-loop erasure probability to the closed-loop erasure probability. Let $p_{ol}(d)$ denote the probability of successful erasure when the open-loop protocol is performed at a duty ratio $d$. Let $p(d,\sigma_n)$ be the probability of successful erasure under feedback erasure protocol. Then,
\small
\begin{align}
  p(d,\sigma_n) = p_{ol}(d) - 0.5\mathbb{P}(z<-\frac{L}{\sqrt{\sigma_T^2+ \sigma_n^2}}),
\end{align}
\normalsize
where, $z = \frac{M_{t_m}-L}{\sqrt{\sigma_T^2+ \sigma_n^2}} \sim \mathcal{N}(0,1),$ and $\mathbb{P}(z<-\frac{L}{\sqrt{\sigma_T^2+ \sigma_n^2}})$ represents the probability of error when a particle is initialized in the right well and with an incorrect decision that the particle is in the left well corresponding to the reset state. To ensure $p$ to be larger than $0.95$ for $d \geq 0.7,$ 
for the values of $\sigma_T$ and $p_{ol}(d)$ (as determined in \cite{talukdar2017memory}), it is determined that $\sigma_n \leq 300 nm$. FIG. \ref{fig:plot1}(d) confirms that $\sigma_n$ should not be more than $300nm$. In this article, $\sigma_n$ is taken to be $300 nm.$ Such a choice allows for the assumption that the Landauer lower bound is a good approximation to the GLB. The mutual information between $X_{t_m}$ and $M_{t_m}$ is given by $I(X_{t_m},M_{t_m}) \approx 0.6.$
\vspace{-8pt}
\subsection{Work Calculations}
\vspace{-5pt}
Stochastic framework for Langevin systems \cite{bhaban2016noise} is utilized to compute the external work by the laser in the feedback process. This quantity depends on the value of $\sigma_n$, as wrong inference about the particle location is a possibility. For an erasure process, the work done on the particle $W_{fb}^d$ is,
\small
\begin{align}
  W_{fb}^d = \begin{cases} 
  0 &\mbox{if } M_{t_m}\leq 0,\\
  \sum_{j=1}^{2}[U(x(t_j),d(t_j^{+}))-U(x(t_j),d(t_j^{-}))] &\mbox{if } M_{t_m}>0,
  \end{cases}
\end{align}
\normalsize
where, $t_1= t_m,\ t_2 = t_f$ are the time instants when the duty ratio is switched from $0.5$ to $d$ and when switched back from $d$ to $0.5,$ respectively. The duration of erasure is $\tau:=t_1 -t_2.$ The choice of $\tau$ is dependent on the value of $d$ for successful erasure (see \cite{talukdar2017memory}). For example, $\tau = 30 s$ for $d=0.7.$

For each memory sample, we computed $W_{fb}^d.$ FIG. \ref{fig:modelFit} shows the average work done by the laser $\langle W_{fb}^d \rangle,$ computed for $d= \{0.7,0.75,0.8,0.85 \}$. We fit the data to the model $\langle W_{fb}^d \rangle_{fit} = A+B\frac{exp(-\frac{0.99}{d-0.5})}{\sqrt{d-0.5}}$ \cite{talukdar2017memory} using Weighted Least Squares. $A,$ represents the minimal average external work required. The value of $A$ is $(0.133 \pm 0.029) k_BT$ and $B$ is $(36.143 \pm 0.958)k_BT.$ $A$ is less than the Landauer limit $k_BT\ln 2$ by a difference of $0.560 k_BT,$ termed as energy deficit, 
and is approximately equal to the mutual information ($\approx 0.6 k_BT$). $d \rightarrow 0.5$ would make the feedback action to approach quasistatic limit. However, there is a practical limitation. Lower $d$ ($<0.7$) would yield in low erasure performance in practice, as evident from FIG. \ref{fig:plot1}(d). Moreover, $\tau$, which is the time chosen for successful erasure is proportional to $\frac{exp[0.99(d-0.5)^{-1}]}{(d-0.5)^{-0.5}}$ \cite{talukdar2017memory} and thus, $\tau $ grows unbounded as $d \rightarrow 0.5$. 
Hence, we extrapolate $\langle W_{fb}^d \rangle$ with the help of the model ($A$) to find the thermodynamic limit ($d\rightarrow 0.5$). 

FIG. \ref{fig:pdf} shows the distribution of $ W_{fb}^d ,$ where the y-axis represents the probability $P(W_{fb}^d).$ 
The spike at origin ($P(W_{fb}^d =0)=0.5$) is almost common for both $d= 0.7$ and $d= 0.75.$ Because, in the feedback step (second step), the potential wells are not tilted when a particle is initialized in the left well and thus the external work done is zero. The lobe present on the right side of the spike contributes significantly to $\langle W_{fb}^d \rangle .$ The right lobe represents the work done in tilting the potentials when a particle is initialized in the right well. By comparing the FIG. $10$ from \cite{talukdar2017memory} with FIG. \ref{fig:pdf}, the difference is a spike present at the origin in FIG. \ref{fig:pdf}.

\vspace{-8pt}
\subsection{Deficit in the Energy Expenditure}
\vspace{-5pt}
We present the thermodynamics of feedback erasure protocol, which would explain the energy deficit observed in the experiments. An average external work $\langle W \rangle$ required in a process is bounded by the information free energy difference $\Delta \mathcal{F}$ (see \cite{deffner2012information}). We now apply second law of thermodynamics to each phase of the protocol.

(a) Measurement: Initial free energy of the system before measurement is $\mathcal{F}(X_{t_m}^{-},M_{t_m}^{-})= \mathcal{F}(X_{t_m}^{-})+\mathcal{F}(M_{t_m}^{-}),$ while the final free energy is $\mathcal{F}(X_{t_m},M_{t_m})= \mathcal{F}(X_{t_m})+\mathcal{F}(M_{t_m})+k_BT I(X_{t_m},M_{t_m})$. The measurement does not perturb the particle position and thus, $\mathcal{F}(X_{t_m}^{-})=\mathcal{F}(X_{t_m})$. From second law, $  \langle W_{meas} \rangle \geq \mathcal{F}(M_{t_m})-\mathcal{F}(M_{t_m}^{-})+k_BT I(X_{t_m},M_{t_m}).$

(b) Feedback: After the measurement, the free energy of the system is $\mathcal{F}(X_{t_m},M_{t_m})$. The feedback action starts at $t_m$ and ends at $t_f,$ and then the system relaxes until $t_e$. The final free energy of the system is $\mathcal{F}(X_{t_e},M_{t_e})= \mathcal{F}(X_{t_e})+\mathcal{F}(M_{t_e}).$ $\mathcal{F}(M_{t_e}) = \mathcal{F}(M_{t_m}),$ as the measurement device is unaffected by the feedback action. From second law, $\langle W_{fb}^d \rangle \geq \mathcal{F}(X_{t_e})-\mathcal{F}(X_{t_m})-k_BT I(X_{t_m},M_{t_m}).$

(c) Reset: The initial free energy of the system is $\mathcal{F}(X_{t_e})+\mathcal{F}(M_{t_e})$ and the final free energy of the system is $\mathcal{F}(X_{t_e}^{+})+\mathcal{F}(M_{t_e}^{+}).$ Reset operation of the measurement device doesn't affect the particle position and hence $\mathcal{F}(X_{t_e}^{+})=\mathcal{F}(X_{t_e})$. Note that $\mathcal{F}(M_{t_e}^{+})=\mathcal{F}(M_{t_m}^{-})$ since the state is reset to its initial value. Replacing $\mathcal{F}(M_{t_e}^{+})$ with $\mathcal{F}(M_{t_m}^{-}),$ $\mathcal{F}(M_{t_e})$ with $\mathcal{F}(M_{t_m}),$ and using second law, we get 
  $\langle W_{res} \rangle \geq \mathcal{F}(M_{t_m}^{-})-\mathcal{F}(M_{t_m}).$

\begin{figure}
	\includegraphics[width=0.85\columnwidth]{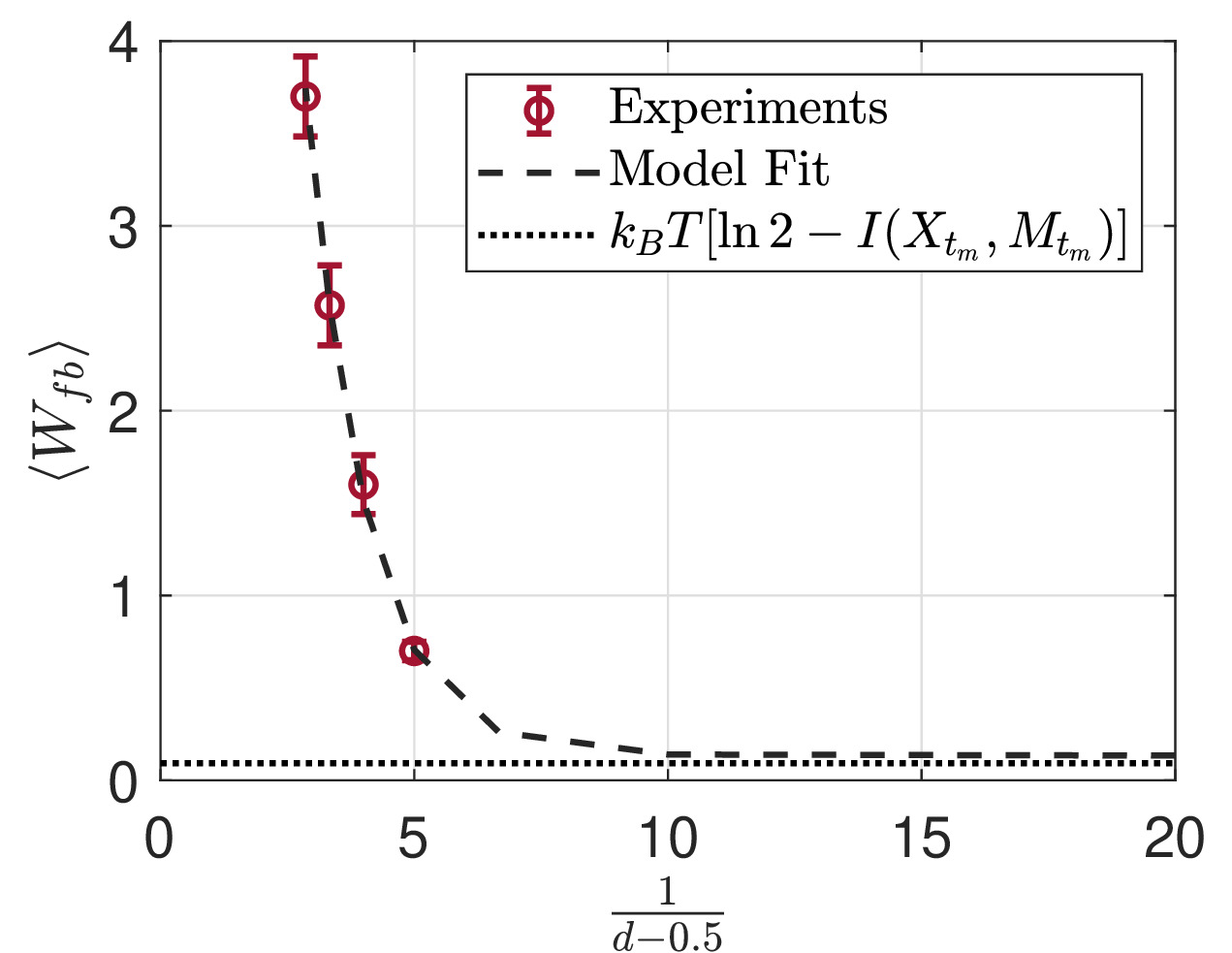}\\
\caption{$W_{fb}^d:$ average feedback work is extrapolated for $d \rightarrow 0.5.$}\label{fig:modelFit}
\end{figure}

The average feedback work $\langle W_{fb}^d \rangle$ consumed by the system is at least $\mathcal{F}(X_{t_e})-\mathcal{F}(X_{t_m})-k_BT I(X_{t_m},M_{t_m});$ and the free energy change associated with system $1$ alone is $k_BT\ln 2$. Thus, the average feedback work is at least $k_BT[\ln 2 -I(X_{t_m},M_{t_m})]$. The difference between the minimal average feedback work and the Landauer limit is $k_BTI(X_{t_m},M_{t_m})$ (mutual information). 
Experimentally we computed $\langle W_{fb}^d \rangle$ and observed to be quasistatically approaching $k_BT[\ln 2 - I(X_{t_m}, M_{t_m})]$ (See FIG. \ref{fig:modelFit}). The energy deficit is given by the mutual information, which was experimentally verified in Section II.C as $0.6 k_BT$. The reduction in energy expenditure is not a violation of second law, as we spend an additional amount of $k_BTI(X_{t_m},M_{t_m})$ in the Measurement. 

\section{Conclusions}
\vspace{-8pt}
A feedback erasure protocol is designed to experimentally realize the lowest average feedback work of $0.58 k_BT,$ smaller than the Landauer limit, for the successful erasure of a memory bit. The feedback action is performed under the influence of noisy measurements of the brownian particle position. To ensure a high erasure probability under the feedback erasure protocol, the measurement noise must be smaller than $300 \ nm.$ The average feedback work for various duty ratio's are computed. We verified that in quasistatic limit, the mutual information ($\approx 0.6 k_BT$) quantitatively explains the deficit in energy expenditure.

\providecommand{\noopsort}[1]{}\providecommand{\singleletter}[1]{#1}%

\end{document}